\documentclass[usegraphicx,usenatbib,useAMS]{mn2e}
\usepackage{times}

\newcommand{\be}{\begin{equation}}
\newcommand{\ba}{\begin{eqnarray}}
\newcommand{\ee}{\end{equation}}
\newcommand{\ea}{\end{eqnarray}}  

\def\lesssim{\mathrel{\hbox{\rlap{\hbox{\lower4pt\hbox{$\sim$}}}\hbox{$<$}}}}
\def\gtrsim{\mathrel{\hbox{\rlap{\hbox{\lower4pt\hbox{$\sim$}}}\hbox{$>$}}}}

\def\gtsima{$\; \buildrel > \over \sim \;$}
\def\ltsima{$\; \buildrel < \over \sim \;$}
\def\gsim{\lower.5ex\hbox{\gtsima}}
\def\lsim{\lower.5ex\hbox{\ltsima}}
\def\simgt{\lower.5ex\hbox{\gtsima}}
\def\simlt{\lower.5ex\hbox{\ltsima}}
\def\simpr{\lower.5ex\hbox{\prosima}}
\def\la{\lsim}
\def\ga{\gsim}

\def\H2{H$_2$}

\def\fH2{f_{\rm H_2}}
\def\j21{J_{21}}

\begin{document}
\title[Fate of clumps in damped Ly$\alpha$ systems]{Fate of
clumps in damped Ly$\alpha$ systems}
\author[I. T. Iliev, et al.]{Ilian T. Iliev,$^{1}$\thanks{e-mail:
iliev@cita.utoronto.ca}
Hiroyuki Hirashita$^{2}$
and Andrea Ferrara$^{3}$
\\
$^1$ Canadian Institute for Theoretical Astrophysics, University of Toronto, 
60 St. George Street, Toronto, ON M5S 3H8, Canada\\
$^2$ {Center for Computational Sciences, University of Tsukuba,
Tsukuba 305-8577, Japan}
 \\
$^3$ SISSA/International School for Advanced Studies, Via
     Beirut 4, 34014 Trieste, Italy \\
}
\twocolumn

\maketitle \label{firstpage}
\begin{abstract}
Recent observations have revealed that damped Ly$\alpha$
clouds (DLAs) host star formation activity. In order to
examine if such star formation activity can be triggered by
ionization fronts, we perform high-resolution hydrodynamics 
and radiative transfer simulations of the effect of radiative 
feedback from propagating ionization fronts on high-density clumps. 
We examine two sources of ultraviolet ({\it UV}) radiation field
to which high-redshift ($z\sim 3$) galaxies could be exposed: one 
corresponding to the {\it UV} radiation originating from stars 
within the DLA, itself, and the other corresponding to the {\it UV} 
background radiation. We find that, for larger clouds, the 
propagating I-fronts created by local stellar sources can trigger 
cooling instability and collapse of significant part, up to 85\%, 
of the cloud, creating conditions for star formation in a timescale 
of a few Myr. The passage of the I-front also triggers collapse of 
smaller clumps (with radii below $\sim 4$ pc), but in these cases the 
resulting cold and dense gas does not reach conditions conducive to 
star formation. Assuming that 85\% of the gas initially in the clump 
is converted into stars, we obtain a star formation rate of 
$\sim 0.25~M_\odot~{\rm yr}^{-1}~{\rm kpc}^{-2}$. This is somewhat higher 
than the value derived from recent observations. 
On the other hand, the background {\it UV} radiation which has harder spectrum
fails to trigger cooling and collapse. Instead, the hard photons 
which have long mean-free-path heat the dense clumps, which as a result
expand and essentially dissolve in the ambient medium. Therefore, the star 
formation activity in DLAs is strongly regulated by the radiative 
feedback, both from the external {\it UV} background and internal stellar sources 
and we predict quiescent evolution of DLAs (not starburst-like evolution).
\end{abstract}

\section{Introduction}

Damped Ly$\alpha$ clouds are quasar (QSO) absorption
line systems whose neutral hydrogen column
density is larger than $\sim 2\times 10^{20}$ cm$^{-2}$
\citep[e.g.][]{prochaska02}. This class of objects dominates
the neutral hydrogen content around $z\sim 3$ and
possibly also at higher and lower redshifts \citep[e.g.][]{peroux03}.
Because QSOs are generally luminous, DLAs provide us with
unique opportunities to trace high-redshift (high-$z$)
galaxy evolution. In particular, by identifying absorption
lines of various species, the physical condition of
the interstellar medium (ISM) of DLAs has been deduced.

{The detection of metal lines in DLAs demonstrates the
existence of heavy elements in DLAs}
\citep[e.g.][]{lu96,ledoux03}. The evolution of metal abundances 
in DLAs can trace the chemical enrichment history of present day 
galaxies. Based on this and other clues, DLAs have been suggested 
to be the progenitors of nearby galaxies; the similar values of the 
baryonic mass density in DLAs around redshift $z\sim 2$ and the stellar 
mass density at $z\sim 0$ has further supported this idea \citep{lanzetta95,SLW00}.
 By adopting a recently favoured Lambda cold dark matter cosmology, however, 
\citet[][]{peroux03} showed that the comoving density of H {\sc i} gas at 
$z\sim 2$ is smaller than the comoving stellar mass
density at $z\sim 0$. Yet they emphasize {that
a large fraction of H {\sc i} gas is contained in DLAs at
$z\sim 2$--3}.

The existence of heavy elements in DLAs indicates that
DLAs have experienced star formation. There is also
direct evidence of { ongoing star formation activity} from the
{\it UV} interstellar radiation field larger than the ultraviolet
background ({\it UVB}) intensity \citep{ledoux02,wolfe03}.
\citet{hirashita03} have theoretically shown that there are small-scale 
clumps whose molecular fraction is higher than $10^{-3}$ in DLAs.
The typical size and density of such clumps are $\sim 5$ pc
and $30$ cm$^{-3}$, respectively. However, it is not
clear if those clouds finally collapse and form stars in
{\it UV} radiation field. It is however crucial to address this question, as most these
clumps are likely to host the most active sites of star formation in DLAs.
\citet{howk05} have shown the existence of
cold gas in a high-$z$ DLA, arguing that such cold
gas may work as the fuel for star formation.

The nature of { DLAs} is at present uncertain. They have been thought
to be the progenitors of nearby large galaxies.
For example, \citet{wolfe86} propose that DLAs are
disk galaxies \citep[see also][]{prochaska98}.
More recently, \citet{haehnelt98} and \citet{ledoux98}
showed that the velocity structure can be explained by infalling 
subgalactic clumps in collapsing dark matter halo. At $z\sim 0$,
spiral galaxies may contribute predominantly to the
damped Ly$\alpha$ absorption \citep{rao93}, but
at high redshift, various classes of objects can be
observed as DLAs \citep{cen02}.
In any scenario, small clumps can form as a result of nonlinear
hydrodynamical processes \citep{wada01}, and it is
important to know if such clumps finally lead to star
formation activity.

In a previous work \citep{hirashita03}, we have used the 
{2-dimensional (2-D)} code of \citet{wada01} 
suitable for disk-like geometry, in order to follow the evolution of
entire galactic disk. { They stress the clumpy structure of
gas, whose typical size is roughly $\sim 5$ pc.}
In this paper, we build on this by examining the 
fate of the resulting clumps in external {\it UV} field.
For this problem, we use a  2-D axially-symmetric radiative-hydrodynamical 
Adaptive Mesh Refinement (AMR) code CORAL
\citep{RTCB95,RML97,MRCLBSN98,SIR04,ISR05}, which is
appropriate for treating each clump.

The structure of this work is as follows.
We describe our numerical methods and input parameters
in \S~\ref{sec:method}. In section \S~\ref{sec:results}
we present and discuss our results and finally, in 
{\S~\ref{sec:summary} we give our summary and
conclusions}.
Throughout the paper 
we use CGS units, unless otherwise noted.

\section{METHOD}
\label{sec:method}

\subsection{Numerical methods}
We use a 2-D axially-symmetric radiative-hydrodynamical 
Adaptive Mesh Refinement (AMR) Eulerian code CORAL 
\citep{RTCB95,RML97,MRCLBSN98,SIR04,ISR05}. The hydrodynamic 
equations are solved by a van Leer flux-splitting method {\citep{vL82}}, 
improved to second order accuracy by use of linear gradients within 
cells as described in \citet{A91}. The microphysical processes 
are coupled to the hydrodynamical evolution by operator-splitting, i.e.
solved at each timestep and results included in the energy equation.
These microphysical processes include photo- and collisional ionizations,
recombinations (both radiative and dielectronic), charge exchange and
transfer of ionizing radiation. The code follows the non-equilibrium
chemical evolution of the ionic species of H, He, { C {\sc ii}--{\sc vi}, 
N {\sc i}--{\sc vi}, O {\sc i}--{\sc vi}, Ne {\sc i}--{\sc vi}, and
S {\sc ii}--{\sc vi} (C~{\sc i} and S~{\sc i} are
assumed fully-ionized)}. Self-gravity is currently not included. The 
refinement (and de-refinement) of a given region in space is decided 
based on the gradients of all code variables, when the gradient is 
larger than a pre-defined value, the region is refined and vice-versa.
We used a total of 8 levels of refinement (separated by factors of 2 
in cell size). 
More details on I-front tracking implementation and detailed tests of 
the method can be found in \citet{SIR04}.

\subsection{\H2 formation and destruction}
\label{subsec:chemi}

For the metallicity range typical of DLAs, we can assume
equilibrium between \H2 formation and destruction, because
the timescale of \H2 formation and destruction is well below
the dynamical timescale \citep{hirashita03}. We adopt the 
formation rate of \H2 per unit volume and time,
$R_{\rm dust}$, by \citet{hollenbach79} 
\citep[see also][]{hirashita03}\footnote{In \citet{hirashita03}
there is a typographic error. (The results are correctly
calculated.) The expression in this paper is correct and $R_1$
in \citet{hirashita03}  is equal to
$R_{\rm dust}nn_{\rm H}(1-\fH2 )$, where $\fH2$ is the
molecular fraction.}:
\begin{eqnarray}
R_{\rm dust} & = & 4.1\times 10^{-17}S_{\rm d}(T)\left(
\frac{a}{0.1~\mu{\rm m}}\right)^{-1}\left(\frac{{\cal D}}{10^{-2}}
\right)\nonumber\\
& \times & \left(\frac{T}{100~{\rm K}}
\right)^{1/2}\left(\frac{\delta}{2~{\rm g}~{\rm cm}^{-3}}\right)
~{\rm cm^{3}~s^{-1}}\, ,
\label{eq:formation}
\end{eqnarray}
where $a$ is the radius of a grain (assumed to be spherical with
a radius of 0.1 $\mu$m in this paper), ${\cal D}$ is the
dust-to-gas mass ratio, $\delta$ is the grain material density (assumed 
to be 2 g cm$^{-3}$ in this paper), and
$S_{\rm d}(T)$ is the sticking coefficient of hydrogen atoms onto
dust. The sticking coefficient is given by \citep{hollenbach79,omukai00}
\begin{eqnarray}
S_{\rm d}(T) & = & [1+0.04(T+T_{\rm d})^{0.5}+2\times 10^{-3}T+8
\times 10^{-6}T^2]^{-1}\nonumber \\
& \times & [1+\exp(7.5\times 10^2(1/75-1/T_{\rm d}))]^{-1}\, ,
\end{eqnarray}
where $T_{\rm d}$ is the dust temperature. We adopt
$T_{\rm d}=20$ K, a typical dust temperature in equilibrium
with the Galactic interstellar radiation field (ISRF) \citep{shibai99}.
However, since the reaction rate is insensitive to the
dust temperature as long as
$T_{\rm d}\la 70$ K, the following results are not affected
by the dust temperature. In fact, some DLAs show ISRF strengths
similar to the Galactic value \citep{ge97,ledoux02,petitjean02}.
The \H2 formation rate per unit volume is estimated
by { $R_1=R_{\rm dust}nn_{\rm H}(1-f_{\rm H_2})$, where $n=1.08n_{\rm H}$
is the total, hydrogen and helium number density, $n_{\rm H}$ is the 
hydrogen number density and
$f_{\rm H_2}\equiv 2n_{\rm H_2}/n_{\rm H}$ ($n_{\rm H_2}$ is the
H$_2$ number density) is the \H2 molecule fraction}.

\begin{figure*}
\includegraphics[width=7.5cm]{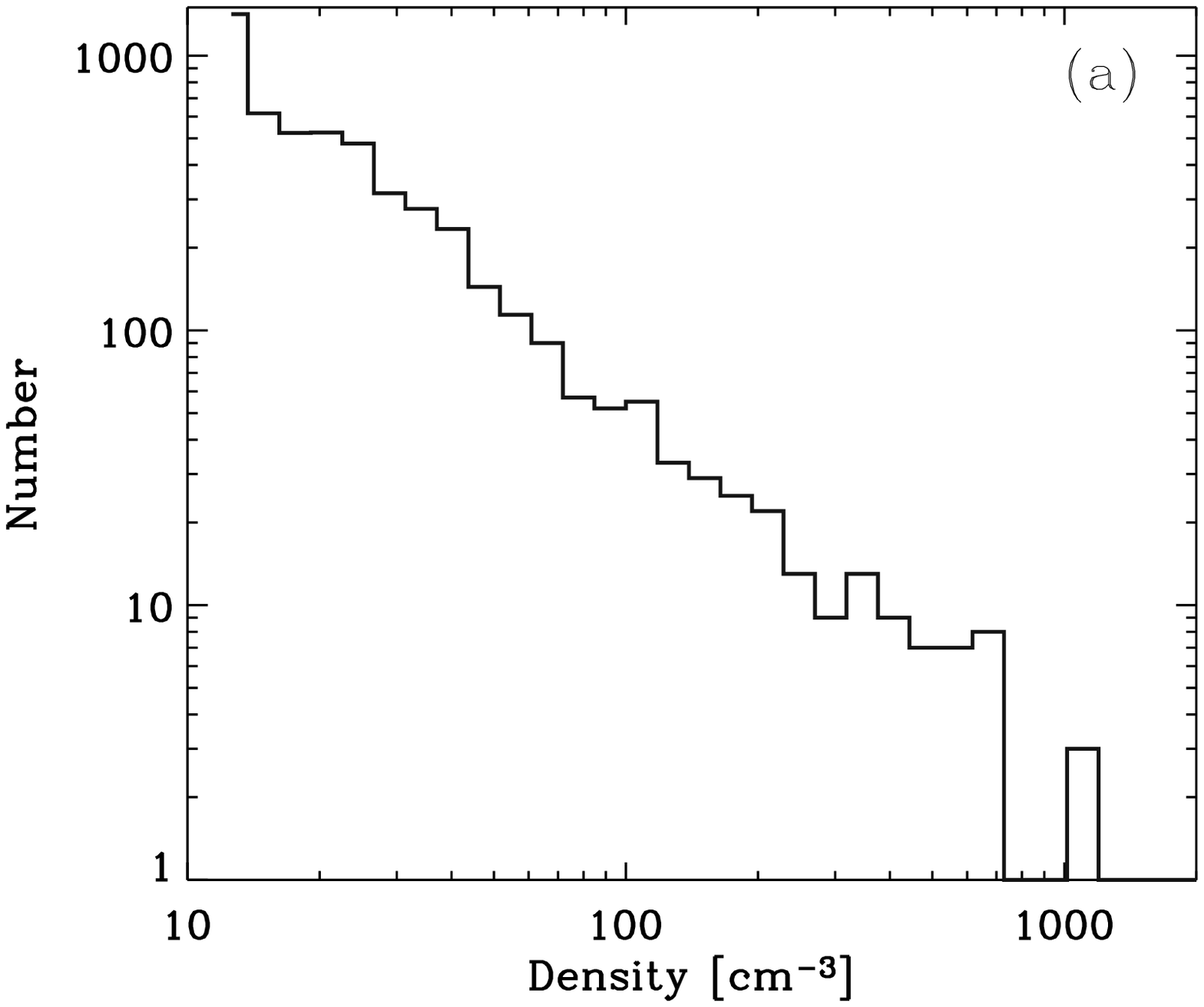}
\includegraphics[width=7.5cm]{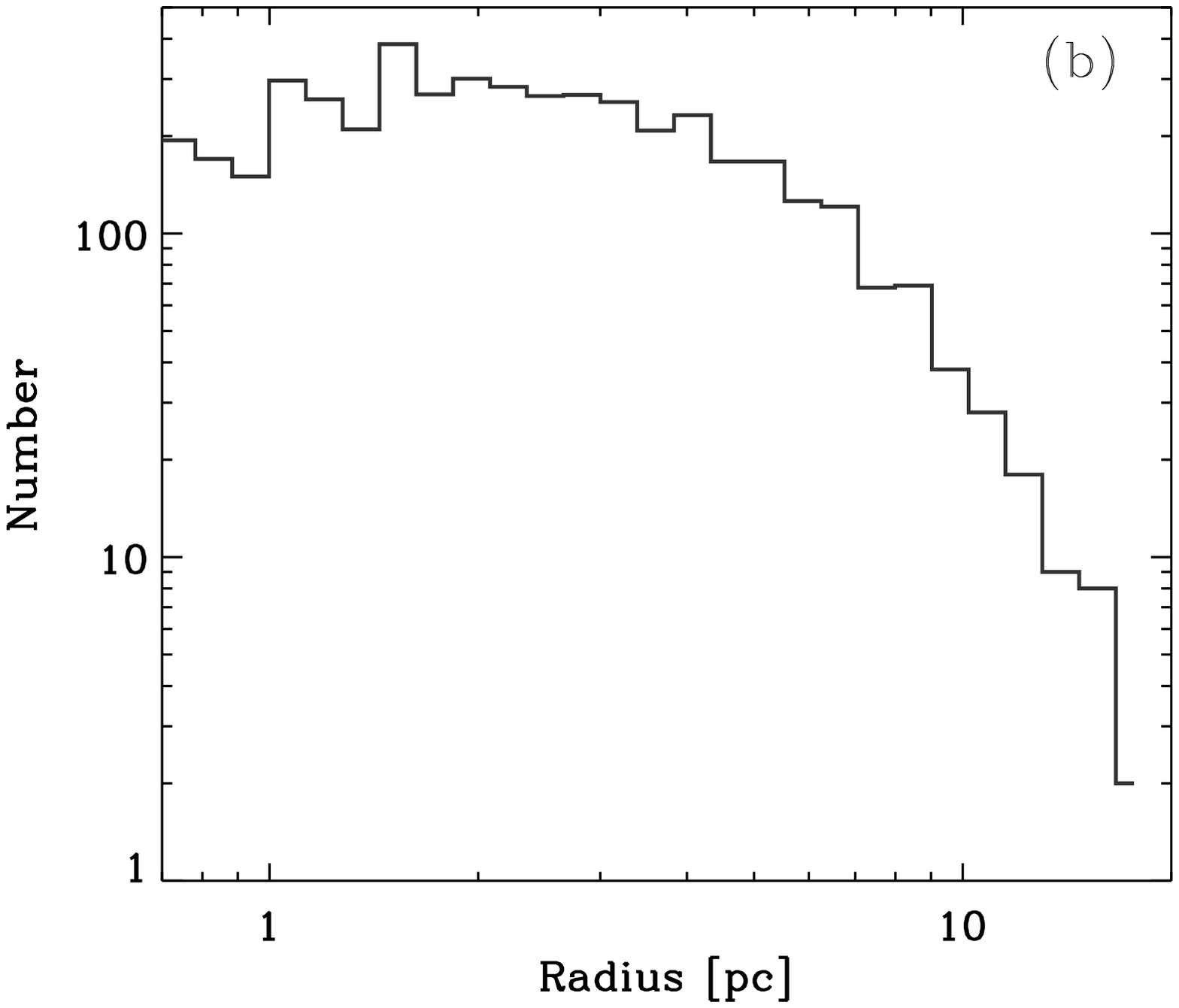}
\caption{Distributions of clumps in disk in (a)(left) mean density and 
(b)(right) size.
\label{clump_distri}}
\end{figure*}

\begin{figure*}
\hspace{-1.4cm}
\includegraphics[width=7cm]{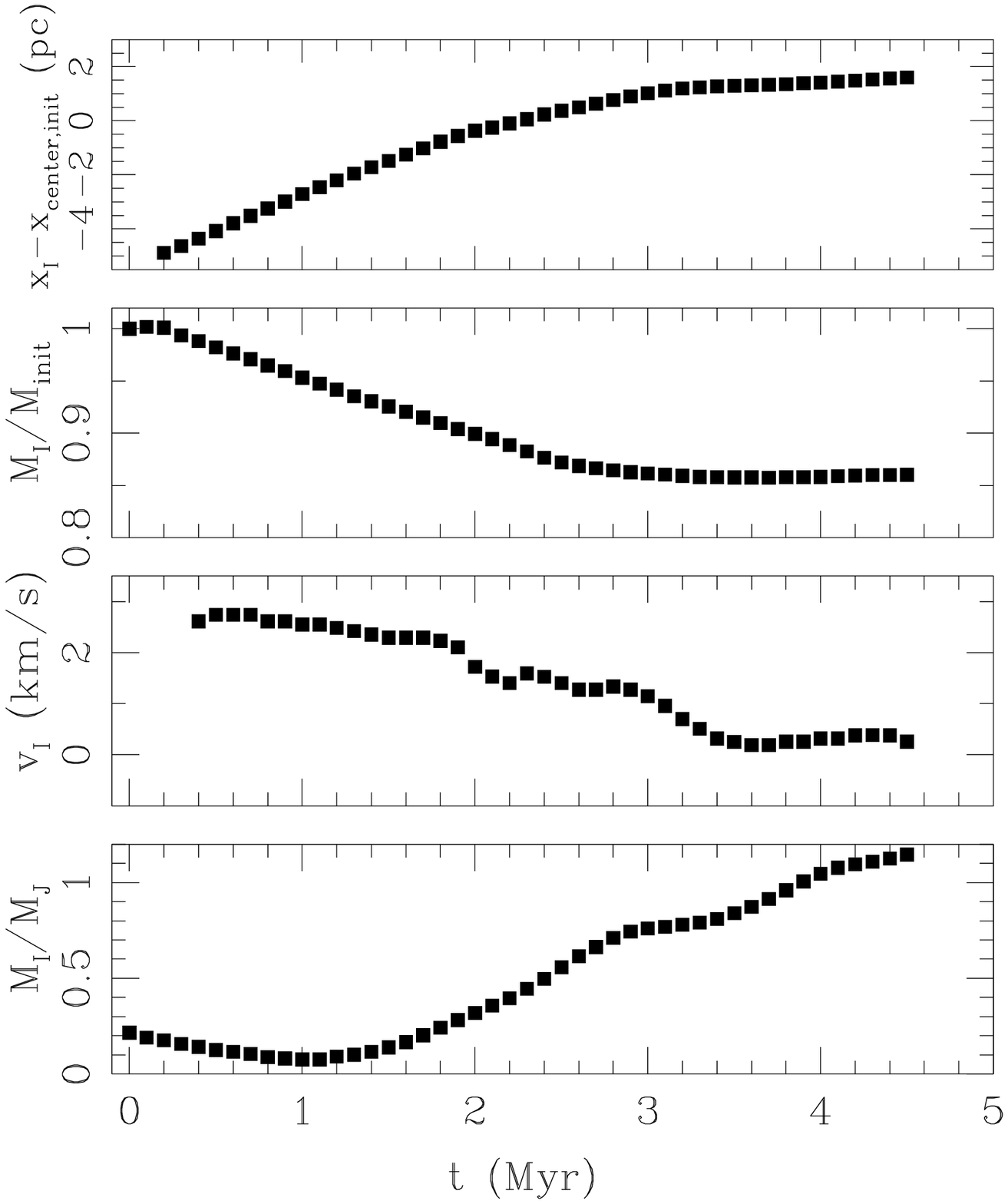}
\hspace{-1.4cm}
\includegraphics[width=7cm]{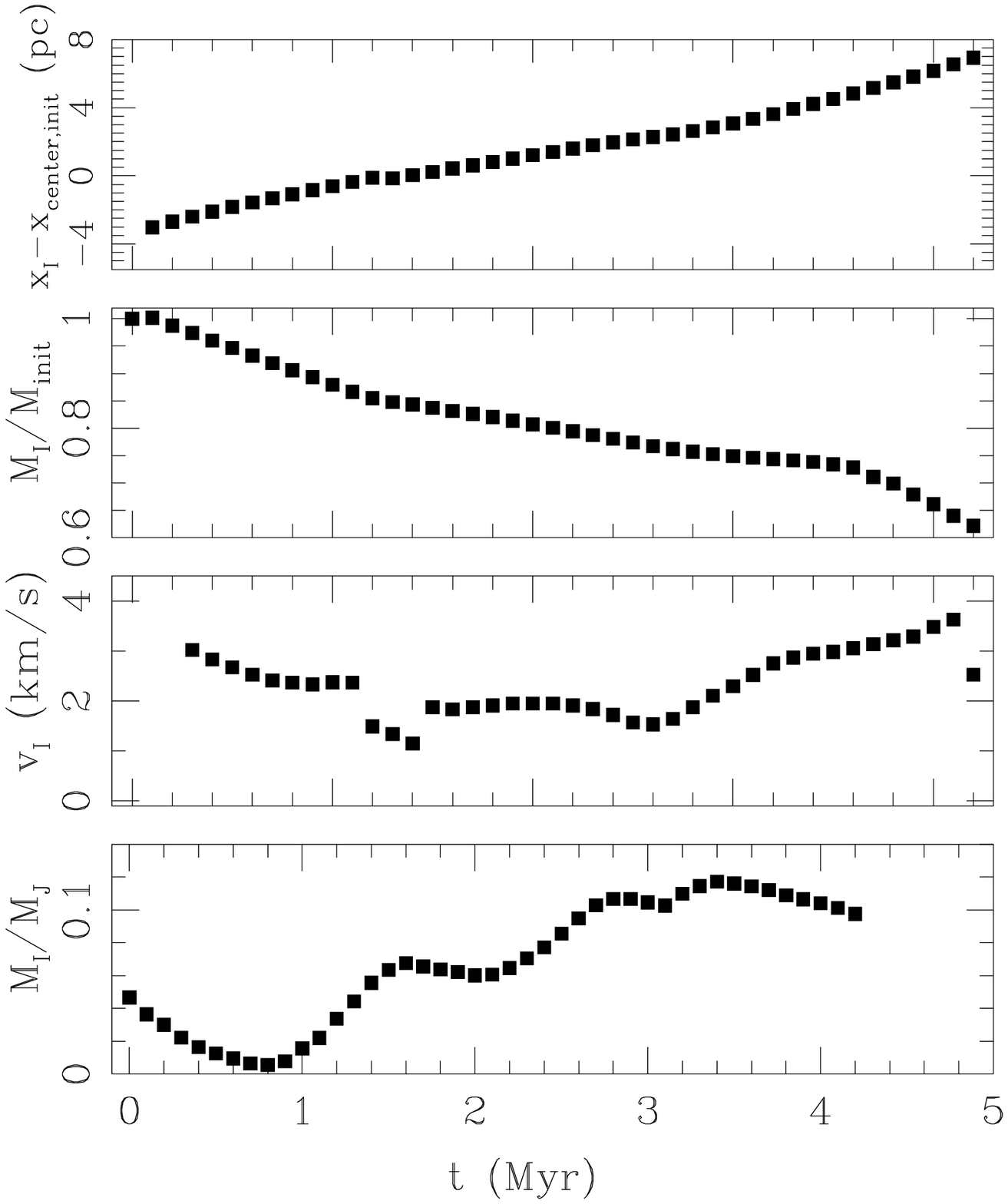}
\hspace{-1.4cm}
\includegraphics[width=7cm]{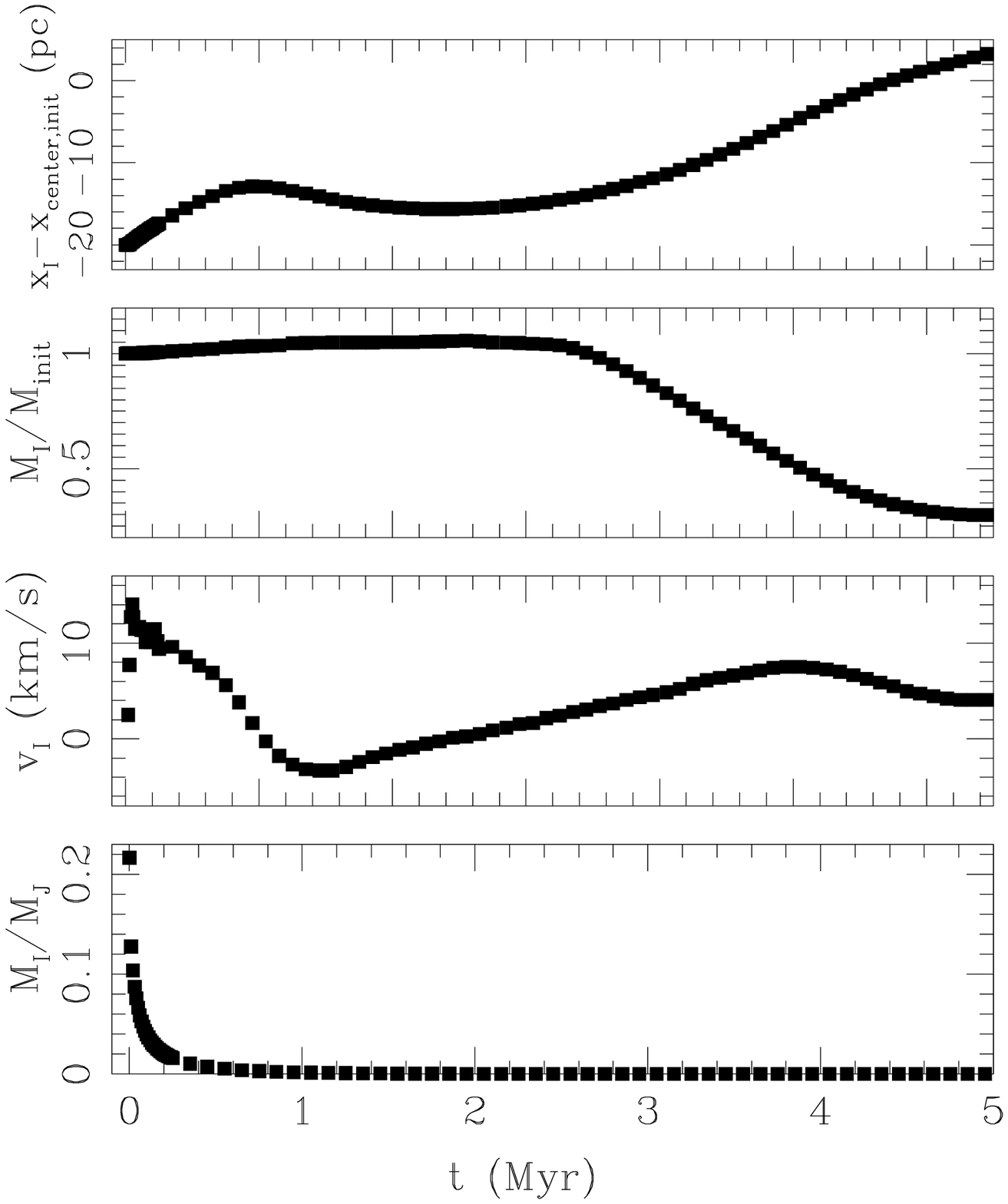}
\caption{Evolution of (top to bottom panels): the position of the I-front along x-axis, 
$x_I$, the neutral mass fraction remaining in the clump, $M_I/M_{\rm init}$, the velocity
of the I-front $v_I$, and the ratio of the cold, neutral mass to the current (averaged)
Jeans mass, $M_J$, for Cases 1 (left), 2 (middle) and 3 (right). 
\label{rocket}}
\end{figure*}

The photodissociation rate $R_{\rm diss}$ is estimated by
\citep{abel97}
\be
R_{\rm diss}=(4\pi )\, 1.1\times 10^{8}J_{\rm LW}\,
S_{\rm shield}~{\rm s}^{-1}\, ,
\label{eq:R_diss}
\ee
where $J_{\rm LW}$ (erg s$^{-1}$ cm$^{-2}$ Hz$^{-1}$
str$^{-1}$) is the {\it UV} intensity at $h\nu =12.87$ eV
averaged for all the solid angle, $S_{\rm shield}$ is the
correction factor of the reaction rate for \H2 self-shielding
and dust extinction. The photodissociation rate per unit
volume is then estimated by $R_2=R_{\rm diss}n_{\rm H_2}$, where
$n_{\rm H_2}$ is the number density of \H2.
We adopt the correction for the \H2
self-shielding by \citet{draine96}(see also
Hirashita \& Ferrara 2002). Then, we estimate
$S_{\rm shield}$ as
\begin{eqnarray}
S_{\rm shield}={\rm min}\left[1,\, \left(
\frac{N_{\rm H_2}}{10^{14}~{\rm cm}^{-2}}\right)^{-0.75}\right] 
e^{-\sigma_{\rm dust}N_{\rm dust}}\,,
\end{eqnarray}
where $N_{\rm H_2}$ is the column density of \H2. The last term,
which is due to absorption by dust is in fact of no relevance for 
the parameters we assume here { \citep{hirashita04}}. 

The metallicity level of DLAs, $\sim 0.1~Z_\odot$ ({\citealt{pettini94}}),
which is the value we assume here, implies that
the dust-to-gas ratio of DLAs is typically 10\% of the
Galactic (Milky Way) value. In this paper, we assume the
Galactic dust-to-gas ratio to be 0.01. We define the normalized
dust-to-gas ratio $\kappa$ as
\begin{eqnarray}
{\cal D}=\kappa{\cal D}_\odot\, ,
\end{eqnarray}
where ${\cal D}_\odot$ is the { Galactic} dust-to-gas ratio
(i.e., ${\cal D}_\odot =0.01$), thus we assume 
$\kappa =0.1$ and hence ${\cal D}=10^{-3}$.

The typical Galactic ISRF intensity has been estimated to
be $c\nu u_\nu =1.2\times 10^{-3}$ erg cm$^{-3}$ at the
wavelength of 1000 \AA\ (i.e., $\nu =3.0\times 10^{15}$ Hz),
where $u_\nu$ is the energy density of photon per unit
frequency \citep{habing68}. Approximating the energy density of
photons at 1000 \AA\ with that at the Lyman-Werner Band,
we obtain $J_{\rm LW}$ at the solar vicinity,
$J_{\rm LW\odot}$, is approximately
$J_{\rm LW\odot}\simeq cu_\nu /4\pi =3.2\times 10^{-20}$ erg
cm$^{-2}$ s$^{-1}$ Hz$^{-1}$ sr$^{-1}$.
The intensity normalized by the Galactic ISRF, $\chi$, is
introduced by
\begin{eqnarray}
J_{\rm LW}\equiv\chi J_{\rm LW\odot}\, .\label{eq:chi}
\end{eqnarray}
Using equations (\ref{eq:R_diss}) and (\ref{eq:chi}),
we obtain
\begin{eqnarray}
R_{\rm diss}=4.4\times 10^{-11}\chi
S_{\rm shield}~{\rm s}^{-1}\, .
\end{eqnarray}
When $\chi =1$, the typical Galactic
photodissociation rate (2--$5\times 10^{-11}~{\rm s}^{-1}$)
derived by \citet{jura74} is recovered.

The {\it UV} background intensity at the Lyman limit is taken  to be
$J_{21}=0.3$--1 around $z\sim 3$, where $J_{\rm 21}$ is
in units of $10^{-21}$ erg cm$^{-2}$ s$^{-1}$ Hz$^{-1}$
sr$^{-1}$ (Giallongo et al.\ 1996;
Cooke, Espey, \& Carswell 1997; Scott et al.\ 2000;
Bianchi, Cristiani, \& Kim 2001). 
For the {\it UVB}, we assume that $J_{21}=0.6$ and that
the spectrum of ionizing radiation is described with
a power law with slope of 1.55. We also take
into account a ``boost'' of $10^{1.5}$ of the
Lyman-Werner photons flux with respect to the ionizing
flux at the threshold, according to the results of
Haardt \& Madau (1996) at $z=2-4$ (i.e.\ we assume
$\chi =0.6$).

The {\it UV} radiation field originating from stars within
DLAs may be larger than the {\it UVB}
(Wolfe et al.\ 2003; Hirashita \& Ferrara 2005).
In particular, the {\it UVB} intensity is likely to be lower
at $z\la 1$ than at $z\sim 3$ (e.g.\ Scott et al.\ 2002).
Therefore, we also examine the case where the {\it UV} field
is dominated by internal stars.
Hirashita \& Ferrara (2005) have derived $\chi =3-30$
by using observational data of H$_2$-detected DLAs.
We adopt $\chi =15$ in this paper.
Although it is known that ionizing photons are
strongly absorbed by the
neutral hydrogen within DLAs, there is little
quantitative constraint on the intensity of ionizing
photons.

However, it would be reasonable to assume that
90\%--99\% (Ciardi, Bianchi, \& Ferrara 2002)
of the ionizing photons are absorbed
relative to the dissociating photons in passing
through the interstellar medium, the ``boost'' could
be 10--100. Assuming the boost of 50, we obtain
$J_{21}=6$ for the internal source.
The spectral shape of the 50,000 K black-body is assumed
for the internal source.

In summary, in our simulations we fix the value of the external ionizing flux 
entering the computational box, $F$, to be equivalent to a mean
isotropic background $J$ of either $J_{21}=6$, with
50,000 K black-body spectrum for the cases where the ionizing sources
and internal, nearby hot stars, or of $J_{21}=0.6$, with power-law spectrum
with slope of 1.55, for the mean intergalactic background (or possibly
internal/nearby QSO sources). The level and spectrum of the
ionizing flux $F(i,j)$ reaching each computational cell $(i,j)$ is 
 determined self-consistently by the code based on the optical depth 
from the source to the cell. Finally, the intensity in the Lyman-Werner 
bands (normalized to the Galactic value), $\chi$, is assumed to be
0.6 for the {\it UVB} and 15 for the internal sources.

 As mentioned at the beginning of this subsection, we can
assume that the formation and destruction of \H2 are
in equilibrium. Therefore, the following equation holds:
\begin{eqnarray}
R_{\rm dust}nn_{\rm H}(1-f_{\rm H_2})=R_{\rm diss}n_{\rm H_2}\, .
\label{eq:equil}
\end{eqnarray}
Using equation~(\ref{eq:equil}), the molecular fraction is
obtained as:
\be
f_{\rm H_2}\equiv\frac{2n_{\rm H_2}}{n_{\rm H}}
    =\frac{2R_{\rm dust}n}{R_{\rm dust}n+R_{\rm diss}}.
\label{fH2}
\ee

\begin{figure*}
\includegraphics[width=8.cm]{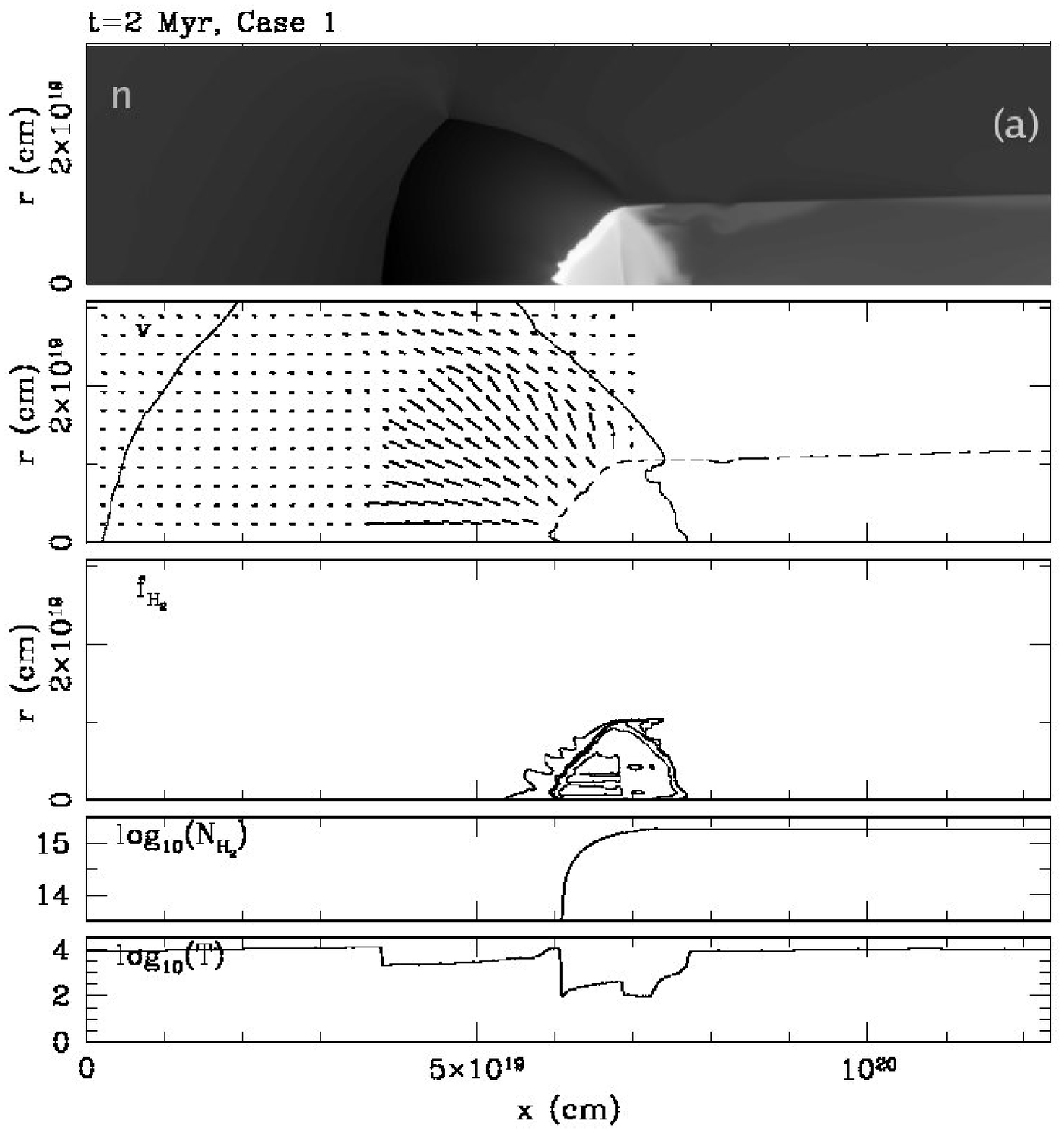}
\includegraphics[width=8.cm]{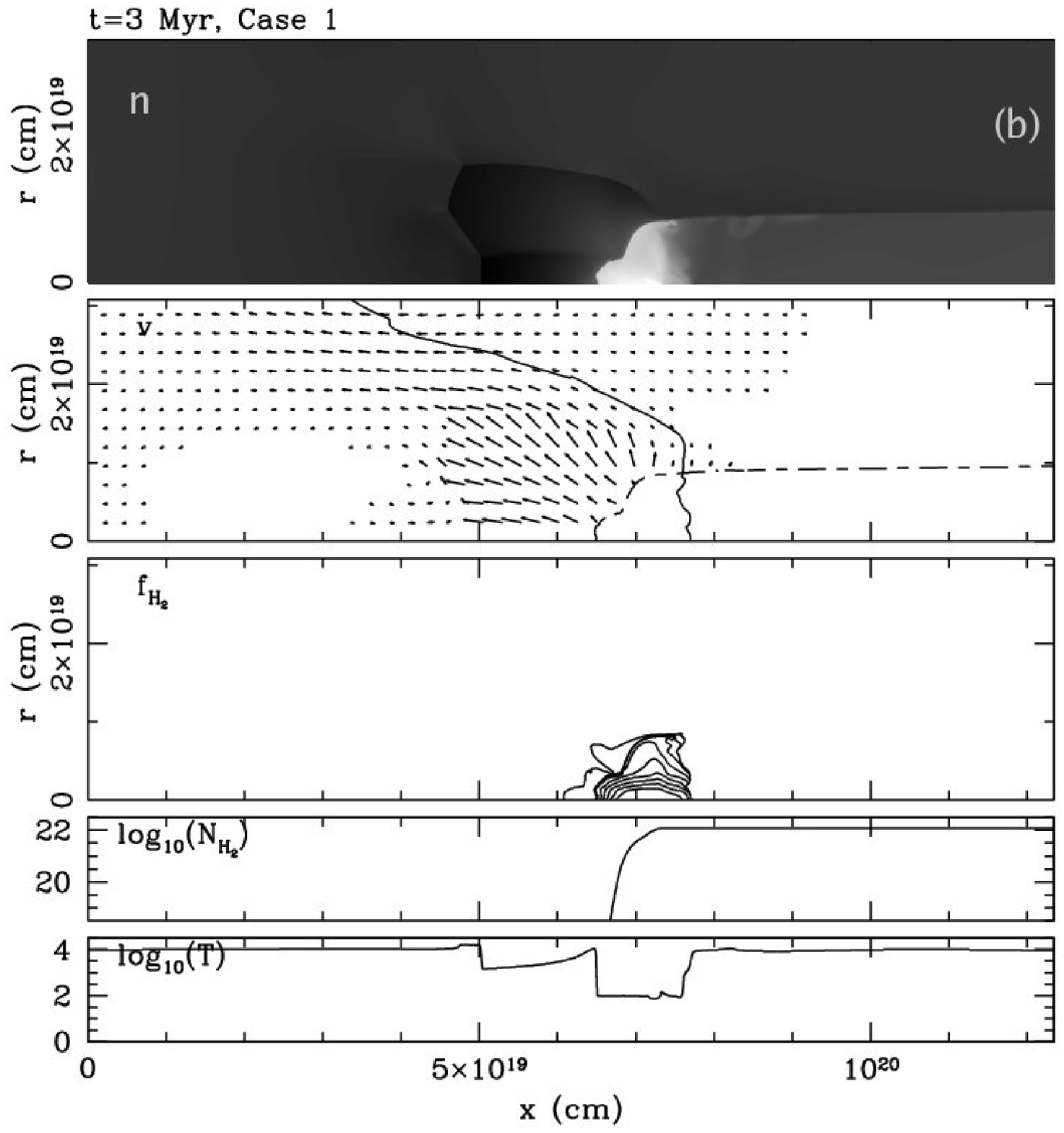}
\includegraphics[width=8.cm]{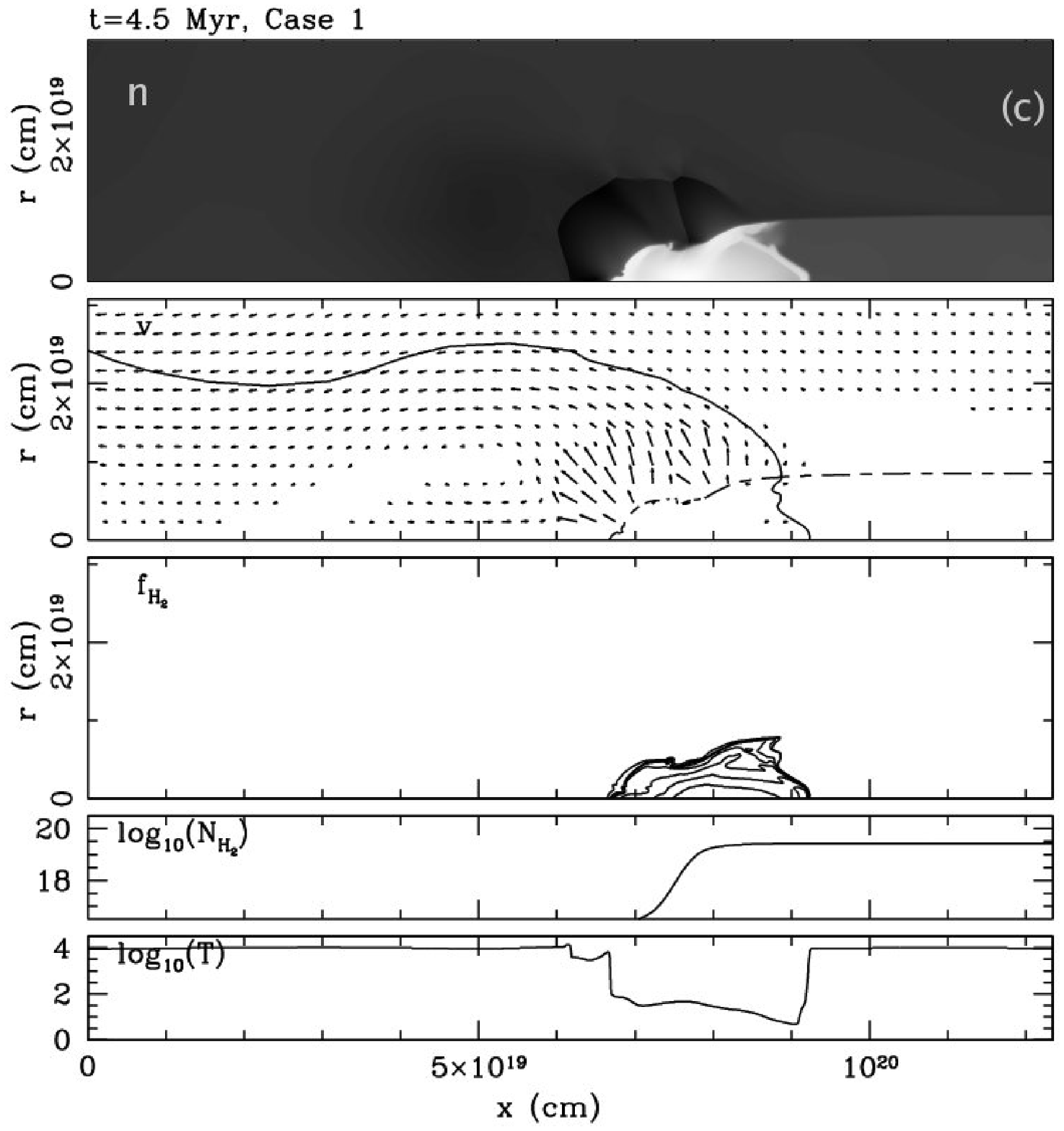}
\caption{Case 1 simulation results at (a) $t=2$ Myrs, (b) $t=3$ Myrs and
(c) $t=4$ Myrs. From top to bottom: (1) grayscale image of the decimal 
logarithm of the atomic density in $(r,x)-$plane of 
cylindrical coordinates; (2) flow velocities; arrows are plotted with length 
proportional to gas velocity. An arrow of length equal to the spacing between 
arrows has velocity $25\, {\rm km\, s^{-1}}$ (i.e. arrows are touching 
along x- and y-axes if $v=25\,\rm km\,s^{-1}$); the minimum velocities plotted are
$\rm 3\,km\,s^{-1}$. Solid line indicates the boundary of the gas which was originally 
inside the cloud, dashed line indicates the current position of the ionization front
(50\% ionization of H);  (3) isocontours of $\rm H_2$ fraction, $f_{\rm H_2}$, 
logarithmically-spaced (from 1 down by factors of 10); cuts along the $r=0$
axis of: (4) molecular hydrogen column density, $N_{H_2}$, ($cm^{-2}$)
and (5) temperature (K).
\label{2-3Myr_case1}}
\end{figure*}

\begin{figure*}
\includegraphics[width=8.cm]{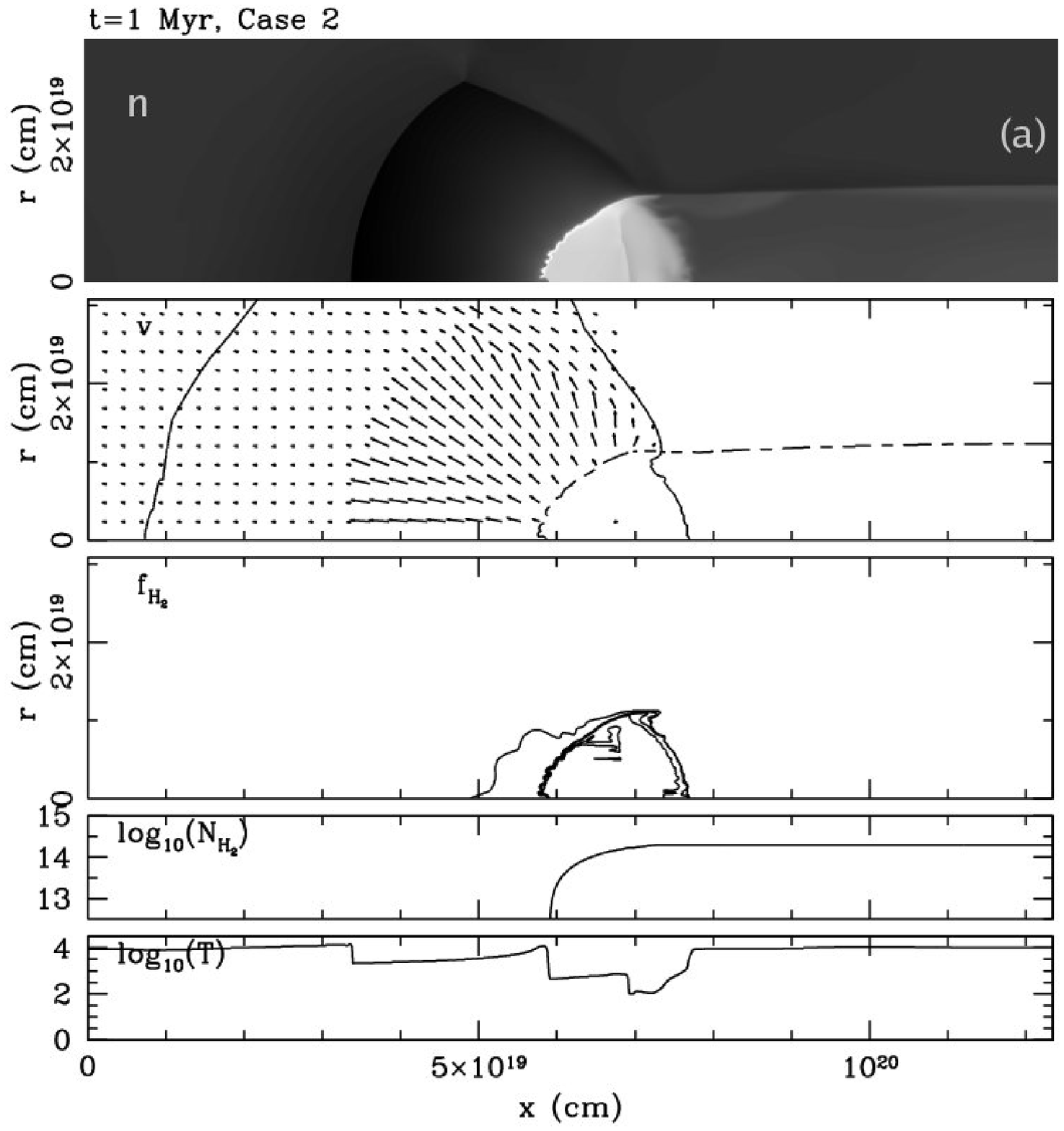}
\includegraphics[width=8.cm]{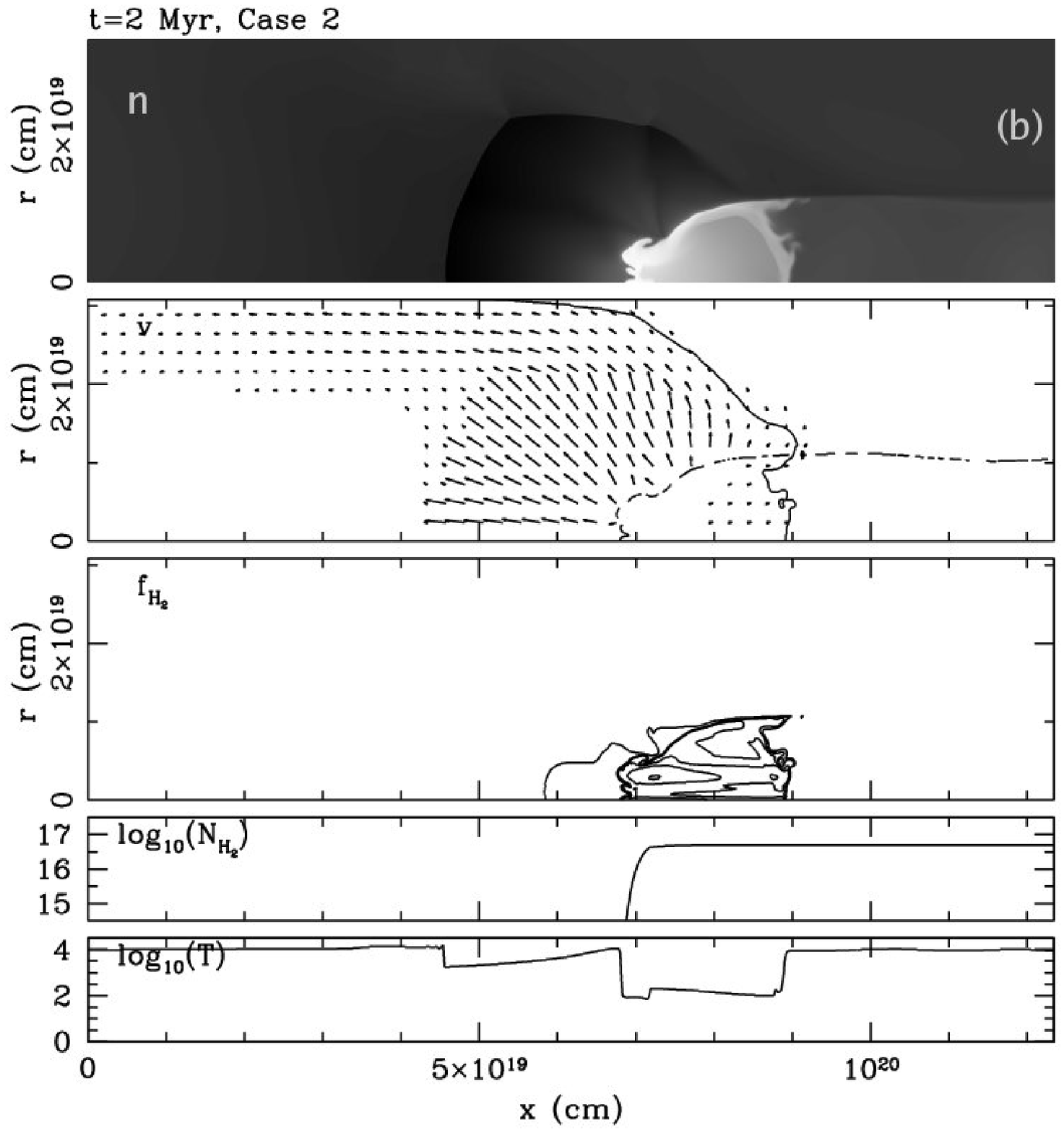}
\includegraphics[width=8.cm]{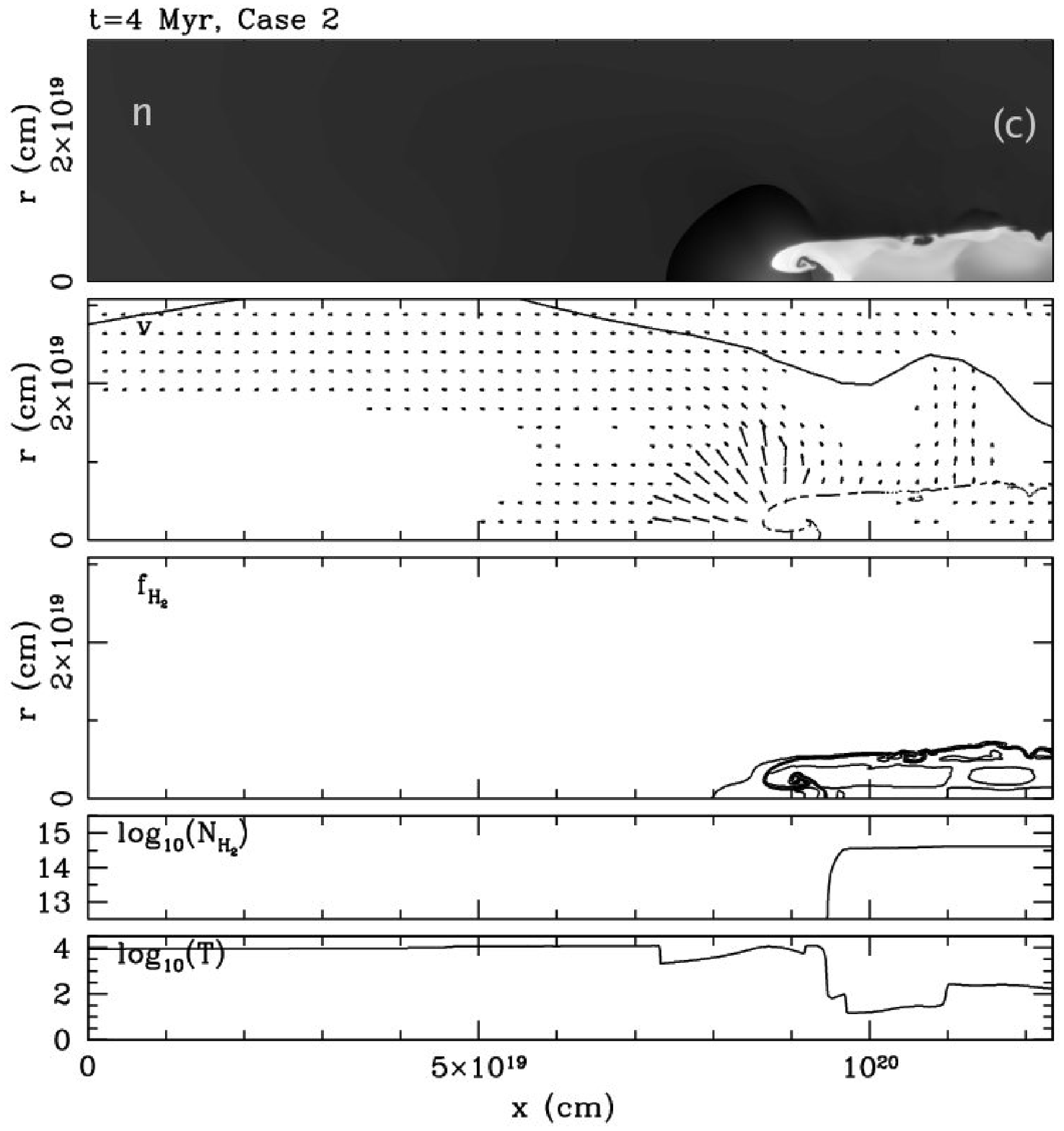}
\caption{Same as in Fig.~\ref{2-3Myr_case1}, but for Case 2 at times (a) 
  $t=$1 Myr, (b) $t=$2 Myr, and (c) $t=$4 Myr.
\label{1-2Myr_case2}}
\end{figure*}

\section{Results}
\label{sec:results}

\begin{table}
\caption{Simulation parameters}
\label{summary}
\begin{tabular}{@{}llll}
Quantity& Case 1& Case 2& Case 3\\ \hline
spectrum& BB 5e4 & BB 5e4 & PL, $\alpha=1.55$\\[2mm]
$r_{\rm clump}$ (pc)&5&3&5\\[2mm]
$n_{\rm clump}$ ($\rm cm^{-3}$)&30&30&30\\[2mm]
$n_{\rm out}$   ($\rm cm^{-3}$)&0.3&0.3&0.3\\[2mm]
$T_{\rm clump}$ (K)&100&100&100\\[2mm]
$T_{\rm out}$ (K)&$10^4$&$10^4$&$10^4$\\[2mm]
$\j21$ &6&6&0.6\\[2mm]
$M_{\rm cloud}$ ($M_\odot$)&474&102&474\\[2mm]
$f_{\rm H_2,init}$ &$10^{-8}$&$10^{-8}$&$10^{-8}$\\[2mm]
\end{tabular}
\end{table}

We have performed three high-resolution simulations with varying cloud sizes 
and external ionizing flux level and spectrum, summarized in Table~\ref{summary}. 
Cloud sizes and densities were chosen similar to the ones 
found on small scales in the larger-scale simulations of a galactic disk
performed by \citet{hirashita03}. In Figure~\ref{clump_distri} we show the  
density and size distribution functions of the clouds from the simulations in 
\citet{hirashita03}, where each cloud is identified as a 2-D region with
local gas density exceeding $10$ cm$^{-3}$ and the cloud radius is 
defined as $(A/\pi)^{1/2}$ ($A$ is the area of the cloud).

In all cases of our simulations listed in Table 1, 
both the clump and the external medium start 
neutral, and the clumps are in a pressure-supported equilibrium. Our fiducial 
case (Case 1) is a spherical cloud of radius $r=5$ pc, temperature $T=100$ K and 
gas number density of $n=30\,{\rm cm^{-3}}$ (corresponding to a total clump mass 
of $M=474\,M_\odot$), and surrounding medium with temperature $T=10^4$ K and gas 
number density of $n=0.3\,{\rm cm^{-3}}$. The external flux has 50,000 K black-body 
spectrum and $\j21=6$. Our Case 2 simulation has all the same initial conditions as 
Case 1, except the clump is smaller, with initial radius of 3 pc (corresponding to a 
total clump mass of $M=102\,M_\odot$). Finally, we considered a case with power-law 
external ionizing flux, $f\propto \nu^{-\alpha}$ with a slope of $\alpha=1.55$ and
  $\j21=0.6$, and all other initial conditions the same as in Case 1.
The Lyman-Werner flux is fixed in all cases by $J_{\rm LW}=474\j21$, as
described in the previous section.

The Inverse Str\"omgren length, $\ell_S$, i.e. the size along a line-of-sight of the 
inverse static Str\"omgren volume from the source side into the clump gas along the 
axis of symmetry, is given by
\be
\ell_S=\frac{F}{\alpha_H^{(2)} n_H^2},
\label{strom}
\ee 
where $\alpha_H^{(2)}$ is the Case B recombination coefficient for hydrogen [see
\citet{SIR04} for details]. Numerically it is equal to 
$\ell_S=(0.005,0.005,0.0005)$ pc in our Cases 1, 2 and 3, respectively, much 
smaller than the corresponding cloud diameter. Therefore the I-front would
always convert to slow, D-type and become trapped by the cloud.

Cases 1 and 2 were run with $1024\times2048$  finest grid resolution, while
Case 3 with $512\times1024$, since simulations which adopt harder ionizing 
spectra converge more readily {\citep{ISR05}},
and thus do not require such high 
resolution. We have also tested our results for convergence by running 
lower-resolution simulation of Case 1, at $512\times1024$ finest-grid resolution. 
Our results show almost no dependence on resolution; thus our conclusions 
remain unchanged and unaffected by resolution concerns.  
 
We show the evolution of the I-front position, velocity, and the neutral mass fraction
remaining in the clump for Cases 1, 2 and 3 in Figure~\ref{rocket}.
We note that there are significant differences in the evolution of these quantities
in the three cases. In Case 1 about 15\% of the cloud mass becomes evaporated within
the first 3 Myr after the arrival of the I-front, and no mass is lost afterward, with 
the cloud retaining the rest, about 85\%, of its initial mass. The I-front initially
propagates into the cloud with $v_I\sim2\,\rm km\,s^{-1}$. After that, its velocity 
continually decreases until the I-front stalls at $t\sim3$ Myr. The cloud itself is
also moving during that time at $\sim 0.1-0.2\rm km\, s^{-1}$, a manifestation of 
the well-known rocket effect, whereby the photoevaporating gas pushes the cloud away 
from the source due to conservation of momentum.
In Case 2 the evolution is initially similar to the one observed in Case 1, with
the cloud losing $\sim 25$\% of its initial mass in the first $3$ Myr due to 
photoevaporation with a subsequent decrease in the mass loss rate. However, in 
this case the outflow still continues afterward, albeit at a low level (the sharp
decrease of the neutral mass fraction after 3.5 Myr is largely due to the cloud 
partially leaving the box, rather than to mass outflow, as would be seen below). 
The clump is again pushed away from the source due to rocket effect, at somewhat 
larger speed ($\sim2\,\rm km s^{-1}$).
Finally, in Case 3 the cloud stays neutral for large part of its evolution,
expanding into the surrounding gas until it eventually becomes mostly ionized
between 3 Myr and 7 Myr after the arrival of the ionization front and disperses.

These cloud collapse/dispersion timescales due to radiative feedback, which are of 
order 3-7 Myr, are similar to the formation/survival timescales obtained 
from the {gas-dynamic simulations in \citet{hirashita03}},
which can be roughly 
estimated as 
$t_{\rm survive}\sim r_{\rm cloud}/v_{\rm flow}\approx 5 \rm pc/1 km\,s^{-1}= 5 Myr$,
where $v_{\rm flow}$ is a typical gas flow velocity observed in the simulations.
In the following, we discuss in more detail each of the Cases. 

\begin{figure*}
\includegraphics[width=8cm]{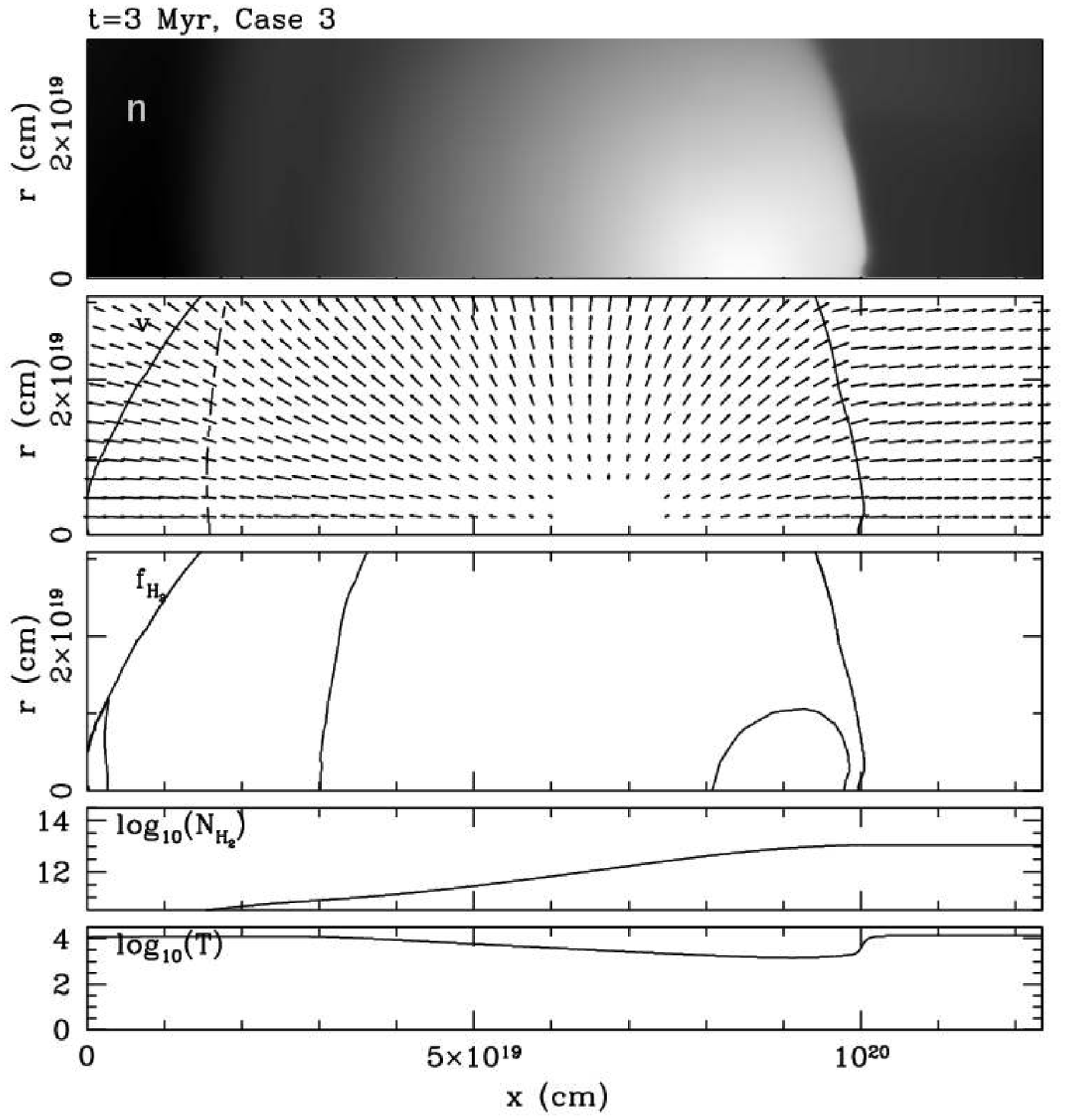}
\caption{Same as in Fig.~\ref{2-3Myr_case1}, but for Case 3 at $t=3$ Myr.
\label{4-3Myr_case2-3}}
\end{figure*}

\subsection{Case 1}
Figure~\ref{2-3Myr_case1} shows snapshots of the gas flow for Case 1 at $t=2$, 
3 and 4.5 Myr. When the I-front reaches 
the clump (at time a fraction of a Myr) it slows down considerably, converts to
D-type and gets trapped. Initially, as the I-front is entering the clump, some 
Raleigh-Taylor instabilities develop at the boundary between the hot gas outside 
and the cold gas inside. The I-front is clearly preceded by a (oblique) shock 
(at the x-axis it is at $x\sim6\times10^{19}$ cm at time $t=2$ Myr), as 
expected for a D-type I-front. A fraction of the cloud material is photoheated 
and evaporated, outflowing in a supersonic wind with speeds 
$\sim25\,{\rm km\,s^{-1}}$. The rest of the clump gas is strongly 
compressed by the propagating shocks. The dense, shocked neutral gas 
has sufficient column density to partially shield the unshocked part of 
the cloud and the hydrogen is already almost fully-molecular $f_{\rm H_2}\sim1$ 
there, while the $\rm H_2$ column density through the cloud, $N_{\rm H_2}$, 
exceeds $10^{19}\,\rm cm^{-2}$, and the molecular hydrogen self-shields. 
By $t=2.5$ Myr 
the right-propagating shock which precedes the I-front crosses the whole clump. 
At $t=3$ Myr (Figure~\ref{2-3Myr_case1}, (b)) the clump is compressed to
less than a third 
of its original radius, and corresponding gas densities $n\simgt10^3\rm cm^{-3}$. By 
this time the neutral portion of the cloud is fully self-shielded
{and the hydrogen is completely molecular ($f_{\rm H_2}\sim1$) there, 
while the column density of H$_2$, $N_{\rm H_2}$,
exceeds $10^{22}\,\rm cm^{-2}$}. By $t=4.5$ Myr the photoevaporation wind
almost disappears, with
the neutral mass of the cloud remaining stable. The clump is again close to uniform
at this time, but compressed to a fraction of its original volume and gas number 
density of $n\sim10^3\rm cm^{-3}$. The clump remains self-shielded, contains 
about 85\% of its original gas, and the hydrogen inside is largely in molecular 
form. The Jeans mass is given by
\be
M_J=\left(\frac{k_B}{Gm_p}\right)^{3/2}\frac{T^{3/2}}{\rho^{1/2}},
\label{M_J}
\ee
where $k_B$ is the Boltzmann constant and $G$ is the gravity constant
and $\rho$ is the gas mass density. In Figure~\ref{rocket} we plot the 
ratio of the mass of the cold, neutral gas remaining to the mean Jeans 
mass at that time for all three simulations. The average Jeans mass is 
calculated based on the mass-weighted temperature and the mean gas density. 
For the compressed cloud ($t\sim 4$ Myr) we have approximately $T\approx100$ K and
$n\approx500\, \rm cm^{-3}$, and $M_J\approx 377\, M_\odot$, while the mass
of the neutral gas in the cloud at this point is about $400 M_\odot$, hence
the cloud should become Jeans-unstable (see Fig.~\ref{rocket}). Later on 
($t\sim4.5$ Myr) the cloud expands and the density drops slightly, but the
temperature also drops due to adiabatic cooling to $T\sim60$ K, and the
local Jeans mass decreases a bit more. The physical conditions reached 
suggest the cloud becomes gravitationally-unstable, cold and molecular, which 
should lead to increased star formation in this case, triggered by the
original I-front. 

\subsection{Case 2}
Results from our Case 2 at $t=1$, 2 and 4 Myr are shown in Figure~\ref{1-2Myr_case2}.
 In this case again the I-front is trapped in 
the cloud and converted to a D-type, preceded by a shock, and expanding
a supersonic wind into the surrounding gas, as is clearly seen in 
Figure~\ref{1-2Myr_case2}. { At $t\sim 1$ Myr, some molecular hydrogen is
created in the pre-shock part of the clump, but its fraction is very small,
 less than $10^{-4}$. $N_{\rm H_2}$ reaches $\sim10^{17.5}\,\rm cm^{-2}$,
and the cloud self-shields}.

The shocks again compress the cloud, similarly to Case 1, but the maximum 
gas number density reached in the course of the evolution is somewhat lower 
than in Case 1, never exceeding $10^3\rm cm^{-3}$ (Figure~\ref{1-2Myr_case2}). 
At $t\sim3$ Myr the clump reaches maximum 
density and the molecular fraction is {$f_{\rm H_2}\sim 1$},
while $N_{\rm H_2}$ 
reaches maximum of $\sim10^{20.5}\,\rm cm^{-2}$, becoming completely self-shielded. 

Afterward, the gas outflow due to photoevaporation continues, albeit
at a somewhat diminished level and the remaining cloud is accelerated.  
As the cloud is pushed back it expands and elongates, and its density 
decreases to $n\sim10^2\rm cm^{-3}$. By $t=4$ Myr much of the molecular 
hydrogen is exposed to the radiation again and mostly destroyed, with
$f_{\rm H_2}$ decreasing to less than $10^{-2}$ everywhere,
while $N_{\rm H_2}$ 
decreases to $\sim10^{18}\,\rm cm^{-2}$, however the cloud is still self-shielded.  

At the point of maximum compression its density and
temperature are $n\sim10^3\,\rm cm^{-3}$ and $T\sim100$ K, respectively, and
the local Jeans mass exceeds the mass of neutral gas remaining in the cloud.
After the cloud expands again ($t\sim4$ Myr) its density drops to 
$n\sim10^2\,\rm cm^{-3}$ and its temperature drops to $T\sim60$ K, in which
case the Jeans mass becomes $M_J\sim 600\,M_\odot$, while the neutral gas remaining
in the cloud is $M\sim70\,M_\odot$, i.e. it never becomes gravitationally-unstable
(see Fig.~\ref{rocket}).
While the cloud is still self-shielded at that time, the hydrogen molecule
fraction is low{:} and conditions are not conducive to star-formation
in this case.

\subsection{Case 3}
The results for Case 3 are illustrated in Figure~\ref{4-3Myr_case2-3}. The
evolution in this case is quite different from the previous two cases.
The I-front is initially trapped outside the clump itself, but the hard
photons which are present in this case reach the cloud, since they have much
longer mean free path and start heating it. The cloud becomes heated to 
$T\sim10^3-10^4$ K and never gets strongly compressed, but instead expands
almost uniformly in all directions due to the higher pressure inside it. It 
eventually
evaporates completely and never forms significant molecular hydrogen fraction.

\subsection{Comparison with observations}
\label{sec:conclusions}

Recent observations have revealed that DLAs host star formation activity 
(e.g., Wolfe et al.\ 2003). 
We have found that propagating I-fronts { produced by} internal stellar sources 
can trigger cooling instability (i.e. a ``compression-cooling-further 
compression'' cycle) and collapse of average clumps (with 
$r\ga 5$~pc and $n\sim 30$ cm$^{-3}$)in DLAs. 
The hydrogen in the collapsed clumps is largely in molecular form and the 
temperature stays below 100 K. The physical conditions are, therefore, 
favourable for star formation. This star formation is triggered by the gas 
compression due to the passage of the I-front. Hence, the {\it UV} produced by 
internal stellar sources works as a ``positive feedback'' for the star 
formation in DLAs. Based on our calculation, we propose that star formation 
in DLAs can trigger further star formation in dense clumps inside the DLA.

When this cooling instability is triggered by the passing I-front, as the cloud 
collapses its {H$_2$} column density can temporarily reach very
high values, up to $10^{22}\rm cm^{-2}$ for clouds with $r\sim 5$~pc. However, 
this phase is quite short-lived 
and the cloud rebounds and re-expands moderately within less than 1 Myr. Due to
this short timescale and the very small cross-section of the clouds, we do not 
expect that such phase would be easily observable. On the other hand, smaller 
clouds, with initial radii $\lesssim4$ pc, also collapse but never
reach { H$_2$} column densities above $\sim10^{18}\rm cm^{-2}$ and most of the 
hydrogen is not in molecular form.

Here we theoretically estimate the star formation rate of a DLA by using the 
cloud data from the simulation by \citet{hirashita03} shown in 
Figure~\ref{clump_distri}. There are $N_A=1092$ clouds of $r>4$ pc per 
each {\rm kpc$^2$} in this simulation, corresponding to mean spacing
between the clouds 
of { $\bar{x}=10^3/N_A^{1/2}=30.3$} pc and mean volume per cloud of 
$\bar{V}=\bar{x}^3=2.8\times10^4\,\rm pc^3$. The total volume of the disk simulated 
in \citet{hirashita03} is 
$V_{\rm tot}=0.785\,\rm kpc^2\times100\,pc=7.9\times10^7\,pc^3$,
corresponding to total number of clouds $N_{\rm tot}=V_{\rm tot}/\bar{V}\approx 2800$.
Assuming a 4 pc cloud, and that 85\% of the gas initially in the clump is converted 
into stars on a time-scale of 3 Myr, as our simulations indicate, this corresponds to 
a star formation rate per cloud 
$\rm SFR_{\rm cl}=0.85M_{\rm cloud}/3\,\rm Myr\approx70\,M_\odot/Myr$,
and total SFR of 
$SFR=SFR_{\rm cloud}N_{\rm tot}\sim 0.25~M_\odot~{\rm yr}^{-1}~{\rm kpc}^{-2}$.
This star formation activity produces a local {\it UV} field. \citet{hirashita04} 
related the surface density of SFR, $\Sigma_{\rm SFR}$, and the {\it UV} field normalized 
to the Galactic value, $\chi$, as follows
\begin{eqnarray}
\Sigma_{\rm SFR}=1.7\times 10^{-3}\chi~M_\odot~
{\rm yr}^{-1}~{\rm kpc}^{-2}\, .
\end{eqnarray}
Therefore, the SFR derived above produces a strong {\it UV} field corresponding to 
$\chi =146$. The SFR estimated above is significantly higher than the value 
derived from recent observations of $0.01-0.05\,M_\odot~{\rm yr}^{-1}~{\rm kpc}^{-2}$ 
\citep{wolfe03,hirashita04}. 
The assumption that all the clouds collapse at the same time could be
unrealistic, and in reality, the star formation activity in DLAs
could be regulated. 
On the other hand, the external {\it UV} background, acts as a 
negative feedback in terms of star formation, as in our Case 3.
Thus, we propose that the star formation activity in 
DLAs is quiescent (i.e. not starburst-like evolution) and strongly regulated 
by a combination of their internal radiative feedback and the external {\it UVB} 
field.

In Cases 1 and 2, the velocity dispersion within the H$_2$-rich regions 
is a few km s$^{-1}$ or less, consistent with the values derived from
H$_2$-detected components \citep{ledoux03}.
The evaporating flow of ionized hydrogen produces a
typical velocity width of $\sim 20$ km s$^{-1}$.
The coexistence of molecular clouds and evaporating
flows may be the reason for the observed complex
velocity structures in H$_2$-detected components
\citep[e.g.][]{petitjean02}.

At the interface of molecular clouds and evaporating flows, H$_2$ suffers {\it UV} 
radiation and it is excited by {\it UV} pumping. This can be the reason for the 
observed high excitation temperature observed for H$_2$ rotational levels $J\ga 2$
(Hirashita \& Ferrara 2004 and references therein). We predict that highly excited 
H$_2$ has a higher velocity dispersion than H$_2$ in low excitation levels,
qualitatively explaining some observations (e.g., an absorption system
at $z=2.595$ in the sight line of Q 0405$-$443; \citealt{ledoux03}).


\section{SUMMARY AND CONCLUSIONS}\label{sec:summary}

We have examined if star formation activity in damped Ly$\alpha$
clouds (DLAs) can be triggered in the presence of {\it UV} radiation.
We have performed high-resolution hydrodynamics 
and radiative transfer simulations of the effect of radiative 
feedback from propagating ionization fronts on high-density clumps. 
We have considered two sources of {\it UV} radiation field
to which high-redshift ($z\sim 3$) galaxies could be exposed: one corresponding 
to the {\it UV} radiation originating from stars within the DLA, itself, and the 
other corresponding to the {\it UV} background radiation. We have found that, for 
clouds with a radius $r\gtrsim4$ pc, the 
propagating I-fronts created by local stellar sources can trigger 
cooling instability and collapse of significant part, up to 85\%, 
of the cloud, creating conditions for star formation in a timescale 
of a few Myr. The passage of the I-front also triggers collapse of 
smaller clumps (with $r\lesssim 4$ pc), but in these cases the 
resulting cold and dense gas does not reach conditions conducive to 
star formation. Assuming that 85\% of the gas initially in the clump 
is converted into stars in clouds with $r>4$ pc, and using the
statistics of the clouds in the simulation of an entire galactic
disk by \citet{hirashita03}, we obtain a surface density of
star formation rate of 
$\sim 0.25~M_\odot~{\rm yr}~{\rm kpc}^{-2}$. This is significantly higher 
than the value derived from recent observations. 
On the other hand, the background {\it UV} radiation which has harder spectrum
fails to trigger cooling and collapse. Instead, the hard photons 
which have long mean-free-path heat the dense clumps, which as a result
expand and essentially dissolve in the ambient medium. Therefore, the star 
formation activity in DLAs is strongly regulated by the radiative 
feedback, both from the external {\it UV} background and internal stellar sources 
and we predict quiescent evolution of DLAs (not starburst-like
evolution). We believe these conclusions are fairly generic and are applicable 
also to molecular clouds in other types of galaxies, as long as the environmental
conditions are similar, in particular in terms of significant presence of metals 
and dust.


\section*{Acknowledgments}
We thank the referee for helpful comments which improved our presentation.
We are very grateful to Alejandro Raga and Garrelt Mellema for allowing us 
to use and modify their code CORAL. This research was partially supported 
by the Research and Training Network "The Physics of the Intergalactic 
Medium" set up by the European Community under the contract 
HPRN-CT2000-00126 RG29185. HH is supported by the University of Tsukuba 
Research Initiative.

\end{document}